%% file: parkcity_chang.tex
\documentclass{edbk} % Computer Modern font calls
\usepackage {epsfig}
\usepackage {colordvi}

\normallatexbib

\newcommand{\lam}{\Lambda}
\newcommand{\lamb}{\overline{\Lambda}}
\newcommand{\pb}{\overline{p}}
\newcommand{\kp}{$K^{+}\:$}
\newcommand{\km}{$K^{-}\:$}
\newcommand{\kmp}{$K^{-}/K^{+}\:$}

\begin{document}

\articletitle[Strangeness Production in Au+Au collisions at AGS by E917]%%  this is the SHORT running title
{Strangeness Production \\
in Au+Au Collisions at the AGS: \\
Recent Results from E917}  %% full title of your chapter

\author{Wen-Chen Chang}

%% affil, email, and abstract are optional
\affil{Department of Physics, University of California, Riverside}

\mbox{}
%\email{wchang@bnl.gov}

\input{e917_collab}

%% optional, to supply a shorter version of the title for the running head:
%%\chaptitlerunninghead{}

\begin{abstract}
Strangeness production in Au+Au collisions has been measured via the
yields of \kp, \km at 6, 8 AGeV and of $\lamb$ at 10.8 AGeV beam kinetic
energy in
experiment E917. By varying the collision centrality and beam energy, a
systematic search for indications of new phenomena and
in-medium effects under high baryon density is undertaken.
\end{abstract}


\section{Introduction}

The study of strangeness production in relativistic heavy-ion
collisions has been of continuing interest as strangeness
is predicted to be enhanced by the formation of a quark gluon plasma
(QGP)\cite{Chang_QGP}. At the same time, many particle properties
such as effective
mass, production threshold, and absorption cross section may be sensitive
to a high baryon density.
Their study might reveal the influence of a many-body mean-field
potential and provide a signal of chiral symmetry restoration.

Experiment E917 at the AGS measured Au+Au collisions at beam kinetic
energies of 6, 8 and 10.8 AGeV in the winter of 1996/97. The Henry Higgins
spectrometer, used previously in experiments E802, E859 and
E866 \cite{Chang_e866kaon}, and an
upgraded data acquisition system enabled the experiment 
to take 280$\times10^6$
kaon-pair/$\pb$ triggered events. The quality of this data set
allows for a detailed study of short lived vector mesons, baryons,
anti-baryons and the systematics of two particle correlations of pion,
kaon and proton pairs. E917 is unique among the AGS experiments in its
ability to measure a wide variety of strangeness-carrying particles
including $K^+$, $K^-$, $\lam$, $\lamb$ and $\phi$-mesons. The study of
the excitation function of kaon production may help identify a possible
phase transition, and the study of $\phi$-mesons may
provide a direct probe of any in-medium effect. More details on the
experimental setup and trigger condition can be found in
Ref.~\cite{Chang_birger,Chang_jamie}.

This article presents a systematic study
of the spectra and yield of \kp and \km in Au+Au collisions at beam
energies of 6 and 8 AGeV combined with published E866
data~\cite{Chang_e866kaon} at 10.8 AGeV. A previously reported
discrepancy in the
measured $\pb$ yields between AGS experiments E878 and E864 has been
hypothesized to arise from an abundance of $\lamb$ production and
different acceptances for the $\pb$ daughter from $\lamb$  decay
in the two experiments.
Experiment E917 is able to make the first direct measurement of
$\lamb$ yields, for which preliminary results are presented.

\section[Kaon Production]{Measurement of Kaons}

There is significant theoretical interest in the study of kaon
properties in dense nuclear matter. Qualitatively, the models
suggest that \kp mesons experience a weak repulsive potential
inside the nuclear medium resulting in a slight increase in the
effective mass with baryon density, whereas
a strong attractive potential for \km mesons leads to a
significantly reduced effective mass in the high baryon density 
environment \cite{Chang_Weise}. Because
the  $\Lambda K^{+}\:$ production channel is expected to be
larger than $K^-K^{+}\:$ pair production at AGS energies,
this scenario results in 
a larger \kmp yield ratio for central events near mid-rapidity, where
high baryon density is expected and the $K^-K^{+}\:$ channel is
enhanced.

The rapidity distributions of kaons were obtained from exponential fits to
the transverse mass, $m_t$, spectra at each rapidity bin. From these
fits we obtain the integrated production probability, $dN/dy$, per unit of
rapidity and the inverse slope, $T_{inv}$, of these spectra. We emphasize
that $T_{inv}$ should not be interpreted as the {\it
temperature} of the emitting source as it is well known that collective
effects, such as radial expansion, can mimic high source temperatures.

\begin{figure}[ht]
\centerline{
\epsfig{file=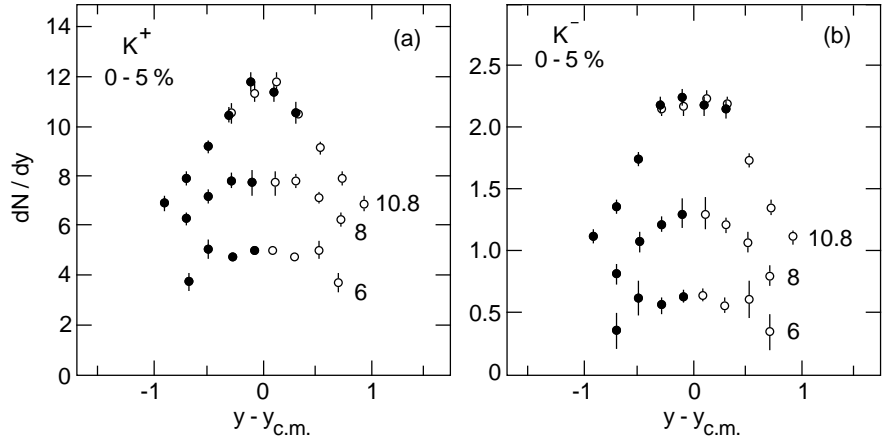,angle=0,width=\textwidth}}
\caption{Rapidity distributions for \kp (panel a) and \km (panel b) 
for 0-5\% central events are shown for beam energies of 6, 8 
(E917 preliminary) and 10.8 AGeV (E866 \cite{Chang_e866kaon}).}
\label{Chang_fig1}
\end{figure}

The rapidity distributions, $dN/dy$, for \kp and \km emission 
for 0-5\% central collisions are shown in Fig.~\ref{Chang_fig1}
for beam energies of 6, 8 and 10.8 ~AGeV . 
The rapidity distributions are observed to be peaked
at mid-rapidity and the yields increase with beam energy without 
substantial change in the shape of the rapidity distributions.
We also find that the rapidity distributions
are essentially independent of centrality.

%%%%%%%%%%%%%%%%%%%%

\begin{figure}[ht]
\centerline{
\epsfig{file=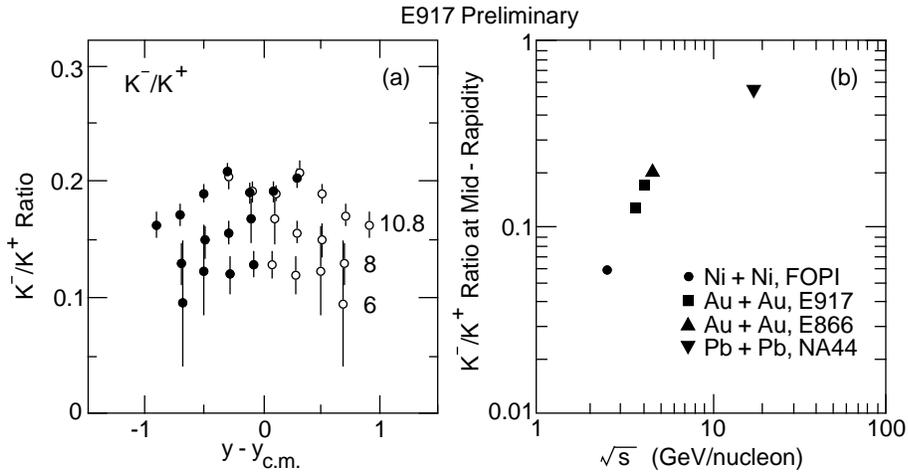,width=\textwidth}}
\caption{ The rapidity distribution of \kmp ratio for the 0-5\%
central events at various beam energies (panel a) and the \kmp ratio at
mid-rapidity at SIS, AGS and SPS~\cite{Chang_fopi_ni, Chang_qm97} (panel b).}
\label{Chang_fig3}
\end{figure}

Since the effects of the nuclear medium have opposite sign for \kp and \km, 
the ratio of \kmp production might be a very sensitive probe for studying 
such effects. It was studied as a function of global parameters for the
collision, such as centrality and rapidity.  In 
Fig.~\ref{Chang_fig3}(a), the \kmp ratio is shown as a function of
rapidity for the 5\% most central events at 6, 8, and
10.8 \cite{Chang_e866kaon} AGeV. 
We observe that the ratio increases with beam energy 
over the rapidity range of this study
and that the rapidity distribution for \km
is narrower than for \kp, an observation that has also been made in studies of
Ni+Ni collisions at SIS energies~\cite{Chang_fopi_ni}. 
The fact that the production of \km relative to
\kp is more abundant around mid-rapidity might be expected, because:
\begin{itemize}
\item the available energy for producing particles is peaked around 
mid-rapidity.
\item the baryon density is largest around mid-rapidity and the in-medium
effect enhances the production of \km relative to that of \kp.
%\item Stronger absorption of \km is expected toward target-beam rapidity.
\end{itemize}
It is, however, difficult to disentangle the relative importance of these two
effects~\cite{Chang_gqli}.

The measurements of the \kmp-ratio at mid-rapidity
is shown in 
Fig.~\ref{Chang_fig3}(b) as a function of center-of-mass
energy from SIS through AGS to SPS energies.
The observed increase in the ratio with beam energy may be expected on the
basis of the higher production threshold for $K^-$. This makes the production
cross section of \km increase faster than that of \kp above the
production threshold~\cite{Chang_gqli}. At SPS energy, the increase in the
ratio is not as steep as that in the lower energy. This is probably
caused by a near saturation in the population of the available phase space for
both \kp and $K^-$.

%%%%%%%%%%%%%%%%%%%%%

We have also studied the centrality dependence 
of \kmp over a wide range of rapidities.
The rapidity distribution of this ratio
exhibits a very weak dependence on centrality at all three energies, 
similar to 
the observations at SIS energies~\cite{Chang_fopi_ni, Chang_kaos_ni}. 
This weak
dependence seemingly contradicts the naive expectation based on the
in-medium effect. Thus, one might expect that the \kmp ratio 
at mid-rapidity should increase
significantly towards central collisions, and that this increase 
should be
most pronounced in the low-energy region close to the production
threshold where the effect of the reduction (increase) of \km (\kp)
mass from the in-medium effect is expected to be strongest. Li and Brown
have proposed~\cite{Chang_gqli} that a suppression
of \km production through the hyperon-feeding channel compensates for 
the increase in the \kmp-ratio from the in-medium effect in the central
collisions. It is, however, difficult to verify this hypothesis 
experimentally.

The inverse slope parameter, $T_{inv}$, derived from the fits to the
$m_t$-spectra is found to peak at mid-rapidity for both \kp and
\km. There is also a slight increase in $T_{inv}$ with beam energy, but the 
rapidity dependence is virtually unchanged.
The difference in $T_{inv}$ for \kp and \km transverse mass spectra 
is found to be small, although the value of $T_{inv}$ for \kp is 
about 50 MeV larger than that of
\km at the  6 AGeV beam energy.

\section{Measurements of $\lamb$ and $\pb$}

AGS experiment E859 has measured a large $\lamb/\pb$ ratio of $2.9 \pm 0.9
\pm 0.5$ for Si+Au at 13.7 AGeV~\cite{Chang_e859}.
This ratio is unexpectedly large
relative to thermal model calculations or with reference to the results for NN
collisions at AGS energies~\cite{Chang_welke}. 
In addition, experiments E864 and E878 at the AGS have 
measured $\pb$ production in the
mid-rapidity region and zero $p_t$ for Au+Pb at 10.6 AGeV~\cite{Chang_e864} 
and Au+Au at 10.8 AGeV~\cite{Chang_e878}
respectively. There is a significant discrepancy between the two experiments
in the reported 
anti-proton production probability for central
events (about a factor of 3.5), but a good agreement for the most
peripheral events. Since these two experiments have different
acceptances for detecting the $\pb$ from $\lamb$ and
$\overline{\Sigma}$ decay, a large production of $\lamb$
might reconcile the results for the two experiments. If this discrepancy 
is  attributed entirely to this effect,
a $\lamb/\pb$ ratio of 3.5  (most probable value) or larger 
than 2.3 (98\% confidence level) is required. 

%%%%%%%%%%%%%%%%%%%%%
\begin{figure}[ht]
\centerline{
\epsfig{file=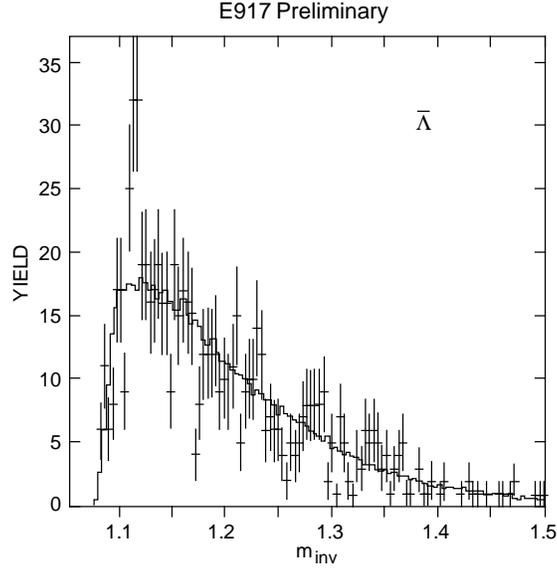,height=8cm}}
\caption{The invariant mass distribution of $\lamb$ reconstructed from
the pair of $\pb$ and $\pi^{+}$. The line is the fitted background
from mixed events.}
\label{Chang_fig6}
\end{figure}

%%%%%%%%%%%%%%%%%%%%%
\begin{table}[hbt]
\caption[The various quantities]{The measurement of $\pb$ and $\lamb$
by E917 in the rapidity interval $y=1.0-1.4$.}
\begin{tabular*}{\textwidth}{@{\extracolsep{\fill}}lcccc} 
\savehline \savehline
           & \multicolumn{2}{c}{Minimum bias} &
\multicolumn{2}{c}{Central events 0-23\%}   \\ \hline
Particle      & $dN/dy$   &  $T_{inv}$  &  $dN/dy$  & $T_{inv}$ \\
              & $(\times 10^{-3})$ & (MeV)  & $(\times 10^{-3}$) & (MeV) \\
\hline
\vspace{5pt}
$\lamb^a$ &$4.3^{+1.8}_{-1.2}$ & $243^{+112}_{-59}$ & $12.9^{+5.5}_{-3.7}$& $243^{+110}_{-60}$\\
\vspace{5pt}
$\pb_{measured}$ & $7.31 \pm 0.17$ & $179 \pm 8$ & $15.0 \pm 0.6$ & $196 \pm 11$   \\
\vspace{5pt}
$\pb_{direct}$    & $4.6^{+0.7}_{-1.2}$ & & $6.8^{+2.3}_{-3.6}$ &\\
\hline
\vspace{5pt}
Ratio $\lamb/\pb_{direct}$& \multicolumn{2}{c}{$0.9^{+0.9}_{-0.3}$} &
\multicolumn{2}{c}{$1.9^{+3.8}_{-0.9}$} \\
\savehline
\end{tabular*}
\begin{tablenotes}
$^a$ dN($\lamb$)/dy = dN($\lamb \rightarrow \pb \pi^+$)/dy / 0.64. \\
%$^b$ dN($\pb_{direct}$)/dy = dN($\pb_{measured}$)/dy - 0.64dN($\lamb$)/dy \\
\end{tablenotes}
\label{Chang_tab1}
\end{table}

\begin{figure}[hb]
\centerline{
\epsfig{file=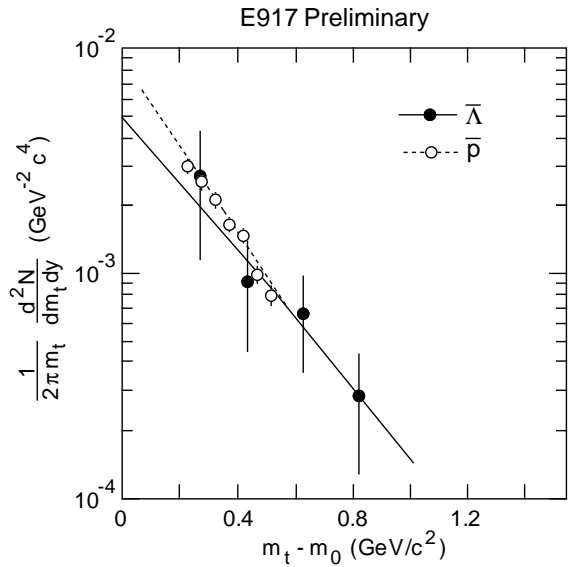,height=8cm}}
\caption{Transverse mass ($m_t$) spectra of $\pb$ (open circles) and
$\lamb$ (solid circles) for the 0-23\% central events.}
\label{Chang_fig7}
\end{figure}

Experiment E917 measured $\pb$ in the rapidity range $1.0<y<1.4$ and
$\lamb$ were reconstructed from $\pb \pi^{+}$ pairs.
The signal of $\lamb$ is clearly
seen in the invariant mass distribution shown in the
Fig.~\ref{Chang_fig6}. The transverse mass spectra of $\pb$ and
$\lamb$ in the rapidity range $1.0<y<1.4$ for the central 0-23\%
events are shown in Fig.~\ref{Chang_fig7}. The efficiency of detecting
$\pb$ from $\lamb$ decay is close to unity in our experiment. Assuming
that the decay of $\lamb$ is the only source of hyperon feed-down into
$\pb$, the yield of $\pb$ directly produced is
$dN(\pb_{\rm direct})/dy = dN(\pb_{\rm measured})/dy-0.64 dN(\lamb)/dy$, 
thereby correcting
for the 64\% branching ratio of the $\lamb$ decay into the $\pb\pi^+$ channel.
The rapidity yield, dN/dy, for $1.0<y<1.4$ and the inverse
slope parameter, $T_{inv}$, obtained from a fit to the $m_t$-spectra
with an exponential function, are listed in Table~\ref{Chang_tab1}. 
Details on this analysis are available in Ref.~\cite{Chang_george}.

%%%%%%%%%%%%%%%%%%%%%

The E917 measurement of the $\lamb/\pb$-ratio is greater than unity for
the 0-23\% central collisions and consistent with the E859 measurement in
Si+Au system. The ratio derived from the difference between the  E864 and
E878 $\pb$ measurements lies within the upper bound of
E917 results. It should be noted, however, that there exist several
differences in the experimental measurements presented here and those
of E864, E878, and E859, as listed in Table~\ref{Chang_tab2}. Most
important are the ranges in $m_t$, rapidity, and centrality measured
by the different experiments.  More data will be analyzed in the
future to compare the results from the other experiments under similar
centrality and rapidity cuts.

\begin{table}[th]
\caption[The difference]{The difference in experimental conditions for
measuring $\lamb/\pb$-ratio.}
\begin{tabular*}{\textwidth}{@{\extracolsep{\fill}}cccccc}
\savehline
\savehline
Exp. & Collision &  $E_{beam}$  & Centrality & Rapidity & $m_t-m_0$ \cr
     &           & (GeV/nucleon)&  (\%)      &          &  (MeV/c$^2$) \\
\savehline
E859 & Si+Au & 13.7 & 0-15 & $1.0-1.4 $ & $> 250$ \cr
E864 & Au+Pb & 10.6 & 0-10 & $1.6-2.0 $ & $= 0$ \cr
E878 & Au+Au & 10.8 & 0-10 & $1.4-2.4 $ & $= 0$ \cr
E917 & Au+Au & 10.8 & 0-23 & $1.0-1.4 $ & $> 250$ \cr
\savehline
\end{tabular*}
\label{Chang_tab2}
\end{table}

\section{Summary and Outlook}

A complete measurement of kaon production at 6 and 8 AGeV has been
presented. No dramatic change is evident in the excitation function of
kaons from 6 to 10.8 AGeV. A straightforward expectation based on the
scenario of many-body in-medium effect on kaons cannot explain the
observation of a weak centrality dependence of \kmp ratios with changing
beam energy.

The $\lamb/\pb$ ratio was measured to be greater than unity for 0-23\%
central collisions. More data need to be analyzed to enable a
detailed comparison with the other results. 

The results presented in this talk are all very preliminary in nature.
For this reason, we have not presented any comparison with, or analysis 
in terms of, theoretical models. 
These will be presented in future publications.

\begin{acknowledgments}
This work was supported by the Department of Energy (USA), the National 
Science Foundation (USA), and KOSEF (Korea).
\end{acknowledgments}

\end{document}

%% file: e917_collab.tex
\small
%\\[\baselineskip]
%
\noindent for the E917 collaboration \\
B.B.~Back$^{1}$, R.R.~Betts$^{1,6}$, H.C.~Britt$^{5}$, J.~Chang$^{3}$,
W.C.~Chang$^{3}$, C.Y.~Chi$^{4}$, Y.Y.~Chu$^{2}$, J.B.~Cumming$^{2}$,
J.C.~Dunlop$^{8}$, W.~Eldredge$^{3}$, S.Y.~Fung$^{3}$, R.~Ganz$^{6,9}$,
E.~Garcia$^{7}$, A.~Gillitzer$^{1,10}$, G.~Heintzelman$^{8}$,
W.F.~Henning$^{1}$, D.J.~Hofman$^{1}$, B.~Holzman$^{1,6}$, J.H.~Kang$^{12}$,
E.J.~Kim$^{12}$, S.Y.~Kim$^{12}$, Y.~Kwon$^{12}$, D.~McLeod$^{6}$,
A.~Mignerey$^{7}$, V.~Nanal$^{1}$, C.~Ogilvie$^{8}$, R.~Pak$^{11}$,
A.~Ruangma$^{7}$, D.~Russ$^{7}$, R.~Seto$^{3}$, P.J.~Stanskas$^{7}$,
G.S.F.~Stephans$^{8}$, H.~Wang$^{3}$, F.L.H.~Wolfs$^{11}$, A.H.~Wuosmaa$^{1}$,
H.~Xiang$^{3}$, G.H.~Xu$^{3}$, H.~Yao$^{8}$, C.M.~Zou$^{3}$
\\[\baselineskip]
$^{1}$ Argonne National Laboratory, Argonne, IL 60439
\\
$^{2}$ Brookhaven National Laboratory, Chemistry Department, Upton, NY 11973
\\
$^{3}$ University of California Riverside, Riverside, CA 92521
\\
$^{4}$ Columbia University, Nevis Laboratories, Irvington, NY 10533
\\
$^{5}$ Department of Energy, Division of Nuclear Physics, Germantown, MD 20874
\\
$^{6}$ University of Illinois at Chicago, Chicago, IL 60607
\\
$^{7}$ University of Maryland, College Park, MD 20742
\\
$^{8}$ Massachusetts Institute of Technology, Cambridge, MA 02139
\\
$^{9}$ Max Planck Institut f\"ur Physik, D-80805 M\"unchen,  Germany
\\
$^{10}$ Technische Universit\"at M\"unchen, D-85748 Garching, Germany
\\
$^{11}$ University of Rochester, Rochester, NY 14627
\\
$^{12}$ Yonsei University, Seoul 120-749, South Korea

\normalsize

%% file: parkcity_chang.bbl
\begin{chapthebibliography}{1}

\bibitem{Chang_QGP}
Koch, P., M\"{u}ller, B., and Rafelski, J. (1986) Phys. Rep. {\bf 142}, 167.
%Shuryak, E. V. (1988) {\it The QCD Vacuum, Hadrons and the Superdense
% Matter}, Part 3 (World Scientific, Singapore.
 
\bibitem{Chang_e866kaon}
Ahle, L. {\it et al.}, (1998) Phys.\ Rev.\ {\bf C58}, 3523.

\bibitem{Chang_birger}
Back, B. (1999) these proceedings.

\bibitem{Chang_jamie} 
Dunlop, J. C. (1999) Ph.D. thesis, MIT.

\bibitem{Chang_Weise}
Weise, W. (1996) Nucl. Phys. {\bf A610}, 35c. 

\bibitem{Chang_fopi_ni} 
Hong, B. (1998) in {\it Proceedings of APCTP Workshop on
Astro-Hadron Physics}, edited by G. E. Brown, World Scientific,
Singapore.

\bibitem{Chang_gqli}
Li, G. Q. and Brown, G. E. (1998) Phys.\ Rev.\ {\bf C58}, 1698.

\bibitem{Chang_qm97}
Bearden, I. G. {\it et al.} (1998) Nucl. Phys. {\bf A638}, 419.

\bibitem{Chang_kaos_ni}
Barth, R. {\it et al.} (1997) Phys.\ Rev.\ Lett. {\bf 78}, 4027.

\bibitem{Chang_e859}
Stephans, G. S. and Wu, Y. (1997) J. Phys. G {\bf G23}, 1895.

\bibitem{Chang_welke}
Wang, G. J.  {\it et al.} (1998) Los Alamos Preprint Archive nucl-th/9807036 
and nucl-th/9806006.

\bibitem{Chang_e864}
Armstrong, T. A.  {\it et al.}, (1997) Phys.\ Rev.\ Lett. {\bf
79}, 3351; Los Alamos Preprint Archive nucl-ex/9811002.

\bibitem{Chang_e878} 
Beavis, D. {\it et al.} (1995) Phys.\ Rev.\ Lett. {\bf 75}, 3633; 
(1997) Phys.\ Rev.\ {\bf C56}, 1521.



\bibitem{Chang_george} 
Heintzelman, G. (1999) Ph.D. thesis, MIT.

\end{chapthebibliography}
